\documentclass{aa}
\usepackage{graphicx}
\usepackage{txfonts}
\begin{document}

   \title{J-PLUS: analysis of the intracluster light in the Coma cluster}

   \author{Y. Jim\'enez-Teja\inst{1,2}\thanks{e-mail: yojite@iaa.es, yolanda@on.br}, 
          R. A. Dupke\inst{1,3,4,2}, R. Lopes de Oliveira\inst{5,6,7,1}, H. S. Xavier\inst{8},  P. R. T. Coelho\inst{8}, A. L. Chies-Santos\inst{9}, C. L\'opez-Sanjuan\inst{10}, A. Alvarez-Candal\inst{1}, M. V. Costa-Duarte\inst{8}, E. Telles\inst{1}, J. A. Hernandez-Jimenez\inst{8}, N. Ben\'itez\inst{2}, J. Alcaniz\inst{1}, 
J. Cenarro\inst{10}, D. Crist\'obal-Hornillos\inst{10}, A. Ederoclite\inst{10}, A. Mar\'in-Franch\inst{10}, C. Mendes de Oliveira\inst{8}, M. Moles\inst{10},  L. Sodr\'e Jr.\inst{8}, J. Varela\inst{10}, \and H. V\'azquez Rami\'o\inst{10}
          }

   \institute{Observat\'orio Nacional, Rua General Jos\'e Cristino, 77 - Bairro Imperial de S\~ao Crist\'ov\~ao, Rio de Janeiro, 20921-400, Brazil\\
              \and
              Instituto de Astrof\'isica de Andalucia, Glorieta de la Astronom\'ia s/n, 18008 Granada, Spain\\
              \and
              Department of Physics and Astronomy, University of Alabama, Box 870324, Tuscaloosa, AL 35487\\ 
              \and
              Department of Astronomy, University of Michigan, 311 West Hall, 1085 South University Ave., Ann Arbor, MI 48109-1107\\
         	  \and 
              X-ray Astrophysics Laboratory, NASA Goddard Space Flight Center, Greenbelt, MD 20771, USA\\
              \and 
              Department of Physics, University of Maryland, Baltimore County, 1000 Hilltop Circle, Baltimore, MD 21250, USA\\
              \and 
              Departamento de F\'isica, Universidade Federal de Sergipe, Av. Marechal Rondon, S/N, 49000-000, S\~ao Crist\'ov\~ao, SE, Brazil\\
              \and 
              Instituto de Astronomia, Geof\'isica e Ci\^encias Atmosf\'ericas, Universidade de S\~ao Paulo, Rua do Mat\~ao 1226, C. Universit\'aria, S\~ao Paulo, 05508-090, Brazil\\
              \and 
              Departamento de Astronomia, Instituto de F\'isica, Universidade Federal do Rio Grande do Sul, Porto Alegre, RS, Brazil\\              
              \and 
              Centro de Estudios de F\'isica del Cosmos de Arag\'on, Unidad Asociada al CSIC, Plaza San Juan 1, 44001 Teruel, Spain\\
             }

   \date{}
  \authorrunning{Jim\'enez-Teja, Dupke, and the J-PLUS collaboration}
  \titlerunning{ICL in the Coma cluster}
 
  \abstract
   {The intracluster light (ICL) is a luminous component of galaxy clusters composed of stars that are gravitationally bound to the cluster potential but do not belong to the individual galaxies. Previous studies of the ICL have shown that its formation and evolution are intimately linked to the evolutionary stage of the cluster. Thus, the analysis of the ICL in the Coma cluster will give insights into the main processes driving the dynamics on this highly complex system.}
   {Using a recently developed technique, we measure the ICL fraction in Coma at several wavelengths, using the J-PLUS unique filter system. The combination of narrow- and broad-band filters provides valuable information on the dynamical state of the cluster, the ICL stellar types, and the morphology of the diffuse light.}
   {We use the Chebyshev-Fourier Intracluster Light Estimator (CICLE) to disentangle the ICL from the light of the galaxies, and to robustly measure the ICL fraction in seven J-PLUS filters.}
   {We obtain the ICL fraction distribution of the Coma cluster at different optical wavelengths, which varies from $\sim 7\%-21\%$, showing the highest values in the narrow band filters {\it J0395, J0410}, and {\it J0430}. This ICL fraction excess is distinctive pattern recently observed in dynamically active clusters (mergers), indicating a higher amount of bluer stars in the ICL compared to the cluster galaxies.}
   {Both the high ICL fractions and the excess in the bluer filters are indicative of a merging state. The presence of younger/lower-metallicity stars the ICL suggests that the main mechanism of ICL formation for the Coma cluster is the stripping of the stars in the outskirts of infalling galaxies and, possibly, the disruption of dwarf galaxies during past/ongoing mergers.}


   \maketitle
%

\section{Introduction}

The Coma cluster ({\it Abell 1656}) is the most massive nearby ($z\sim0.023$) cluster to us, with a virial mass of $M_{200}=1.88\times 10^{15} h^{-1}M_\odot$ \citep{kubo2007}. Originally thought as a classical example of a virialized cluster given its apparent spherical symmetry, compactness, luminosity segregation, and regularity \citep{zwicky1957,kent1982,mellier1988}, Coma now is known to be dynamically very active. It has two D galaxies near the center (NGC 4874 and NGC 4889) and its spatial distribution shows several subclumps and overdensities \citep{fitchett1987,mellier1988,ulmer1994,merritt1994,conselice1998,mendelin2017}. Its galaxy velocity distribution presents evidence for substructures and departure from a single gaussian, such as significant skewness and the presence of several peaks \citep{fitchett1987,merritt1987,kent1982,mellier1988,colless1996}. Many substructures are also found in the elongated X-ray emission of Coma (whose peak is not coincident with any of the BCGs), with a specially relevant arc-like emission between the Coma core and the NGC 4839 group consistent with a bow shock \citep{johnson1979,briel1992,davis1993,white1993,vikhlinin1997,arnaud2001,briel2001,neumann2001,neumann2003}. Coma has a high intracluster gas temperature ($k$T\,$>$\,8\,keV) and lacks significant temperature and metallicity gradients in X-rays within an angular radius of 30 arcmin ($\sim$840 $h^{-1}_{70}$ kpc) from the cluster core  \citep{edge1992,watt1992,colless1996,arnaud2001,simionescu2013}. It also hosts a giant radio halo and a peripheral radio relic connected through a low-surface brightness ridge of emission \citep{willson1970,jaffe1976,giovannini1985,kim1990,venturi1990,deiss1997,kronberg2007,brown2011}. \\ 

A formation scenario that is currently contemplated for the Coma cluster is a merger of two sub-clusters or groups associated with the two D galaxies, NGC 4874 and NGC 4889, possibly after their second core crossing \citep{briel1992,watt1992,davis1993,colless1996,arnaud2001,neumann2003,adami2005b,gerhard2007}. Most dynamical and X-ray studies suggest that the group related to NGC 4874 could have been at the center of the main cluster potential in the past, being displaced  out of the bottom of the potential well by the infall of the group associated with NGC 4889 \citep[e.g.][]{davis1993,colless1996,gerhard2007}. Moreover, analyses of X-ray substrutures and radio emission indicate that this second subcluster is partially disrupted and have ejected NGC 4889 \citep{colless1996}. Evidence also points to the presence of a third subcluster around NGC 4839 that would be currently merging into the main Coma cluster \citep{briel1992,briel2001,colless1996,neumann2001,neumann2003,adami2005b}.   \\
  
\cite{zwicky1957} was the first to report the existence of a diffuse intergalactic light in the core of the Coma cluster, detected by direct photography, to be later confirmed by \cite{gunn1969} using photoelectric drift-scan observations. However, it was not until 1970 that this emission was first measured by \cite{devaucouleurs1970}, concluding that it could represent a 40\% of the integrated luminosity of the bright cluster galaxies in the B-band. These authors were not able to identify the diffuse light in Coma as an independent component. Instead, they associated the ICL to the overlapping halos of the two D galaxies located in the central region of Coma. One year later, \cite{welch1971} studied in more detail the excess of light in the Coma core, finding that it was distributed not only in the surroundings of the two giant ellipticals, but also extended outwards to the southwest from NGC 4874. That piece of evidence confirmed that the stars in the diffuse light might be indeed intergalactic, in particular those from the southwest luminous ridge that was interpreted as matter flowing out from NGC 4874. \cite{kormendy1974} also observed  some luminosity eastward of NGC 4889. The distribution of the ICL in Coma has been confirmed by several authors ever since \citep[e.g.][]{melnick1977,thuan1977}. Comparing the B-V color of the diffuse light with that of the galaxies, \cite{thuan1977} found, surprisingly, the ICL to be bluer at larger radii. The radial color gradient of the stellar populations in the galaxies and the similarity of the color of the ICL in Coma with that of M87 triggered the hypothesis that this diffuse light could be primarily formed by stars tidally stripped from galaxy collisions in the cluster core. The discovery of low surface brightness tidal features in the ICL supported this idea \citep{trentham1998,gregg1998,adami2005}.\\

The ICL fraction, defined as the ratio between the ICL and total cluster's luminosities, has been extensively studied in the Coma cluster. \cite{melnick1977} established an upper limit of 25\% for the ICL fraction in Coma, estimated from their measurements in the Gunn {\it g} and {\it r} bands ($16\pm8\%$ and $19\pm7\%$, respectively).\cite{thuan1977} estimated that the contribution of the ICL to the total luminosity of the cluster in the G-band (4600-5400\AA) could be as high as $\sim 31\%$. Other works measured the ratio between the ICL and the total galaxy luminosities (i.e., the total cluster luminosity without including the ICL), obtaining a 30\% and 56\% in the {\it V} and {\it B} bands, respectively \citep{mattila1977}, a 45\% in the G band \citep{thuan1977}, and a 50\% in the {\it R} filter \citep{bernstein1995}. An exhaustive analysis of the ICL distribution that revealed numerous concentrations of diffuse light in the cluster core was later performed by \cite{adami2005}, who derived an ICL fraction of 20\% in the R-band. We must remark the precision and consistency of all these works, given that they were mainly based on photographic plates (with different levels of sensitivity and depth), being \cite{bernstein1995} the first to use CCD images.\\ 

In this work we aim to measure the diffuse light in Coma using a set of several optical narrow and broad filters, and to determine the clusters dynamical stage through the ICL fraction. In \cite{jimenez-teja2018} we observed that merging clusters showed a distinctive gradient in the ICL fraction measured at different wavelengths, with an excess in the band that comprised the emission peaks of the stars with spectral type near A and F, compared to the nearly constant ICL fractions displayed by relaxed systems. A multi-wavelength analysis of the ICL fraction in the Coma cluster could provide an independent test for its disturbed stage, solely based on optical data. Moreover, the use of narrow-band filters could give us insights on the nature of the different features and substructures observed in the diffuse light as well as the main processes driving its dynamics. \\

Data used for this work comes from the Javalambre-Photometric Local Universe Survey \citep[J-PLUS,][]{cenarro2018}, a multi-wavelength survey built to complement and support the future Javalambre-PAU Astrophysical Survey \citep[J-PAS,][]{benitez2014}, providing technical and scientific knowledge that will be used, among other things, for photometric calibration of J-PAS. As it will be shown later, the J-PLUS filter system, which combines narrow and broad-band filters, is excellent to obtain a detailed characterization of the ICL fraction gradient and to analyze the nature of the stellar populations of the ICL.\\

This work is organized as follows. We describe in Sect. \ref{sect_data} the main characteristics of the images and spectra used. The algorithm applied, called Chebyshev-Fourier Intracluster Light Estimator, is outlined in Sect. \ref{sect_methodology}, while the results obtained are shown in Sect. \ref{sect_results}. The main implications extracted from these results are discussed in Sect. \ref{sect_discussion}, while the final conclusions are summarized in Sect. \ref{sect_conclusions}. Throughout this paper we will assume a standard $\Lambda$CDM cosmology with $H_0$=70 km s$^{-1}$ Mpc$^{-1}$, $\Omega_m$=0.3, and $\Omega_{\Lambda}$=0.7. All the magnitudes are referred to the AB system.\\

\section{Data} \label{sect_data}

J-PLUS data are collected with the 83cm-diameter telescope (JAST/T80) located at the {\it Observatorio Astrof\'isico de Javalambre} (OAJ) in Teruel (Spain). JAST/T80 is provided with the panoramic camera T80cam, which has a field-of-view of 3 deg$^2$ and a sampling rate of $0.55^{\prime\prime}$ per pixel, and a 12 filter system spanning the optical range, from 3500 to 10000 \AA. This set comprises five broad-band ({\it u, g, r, i}, and {\it z}) and seven narrow-band filters ({\it J0378, J0395, J0410, J0430, J0515, J0660}, and {\it J0861}), specially designed to identify and characterize the spectral energy distribution of stars and nearby galaxies. J-PLUS images of the Coma cluster were collected as part of the regular science commissioning program of the JAST/T80, between March 4 and April 14, 2016.  Five narrow- and five broad-band filters were used, as described in Table \ref{tabla_observaciones_resultados}, where the number of exposures and the exposure time for each filter are also indicated.\\

\begin{table*}
\caption{J-PLUS observations of Coma and CICLE results. Observations in three filters were discarded since they were affected by different artifacts that could bias our ICL measurements. The obtained ICL fractions after applying CICLE and the equivalent radii up to which the ICL was found to extend for each filter are shown.}
\label{tabla_observaciones_resultados}   
\centering                    
\begin{tabular}{c c c c c}        
\hline\hline                 
Filter & \# Exposures & Total exposure time & ICL fraction & Radius\\ 
 && (s) & (\%) & (kpc)\\
\hline                        
   {\it u} & 2 & 934 & ... & ...\\
   {\it J0378} & 1 & 1135 & ... & ...\\
   {\it J0395} & 4 & 1844 & $21.0\pm4.1$ & 353\\
   {\it J0410} & 1 & 750 & $16.2\pm12.0$ & 354\\
   {\it J0430} & 2 & 1540 & $20.1\pm13.8$ & 403\\
   {\it g} & 2 & 252 & $7.6\pm3.7$ & 249\\\
   {\it J0515} & 1 & 1035 & $10.1\pm8.3$ & 370\\
   {\it r} & 2 & 490 & $12.3\pm6.0$ & 470\\
   {\it i} & 2 & 978 & $16.1\pm16.3$ & 453\\
   {\it z} & 1 & 3185 & ... & ...\\
\hline                        
\end{tabular}
\end{table*}

Images are firstly reduced using the processing pipeline, {\it jype} \citep{cristobal-hornillos2012}, developed and provided by the Centro de Estudios de F\'isica del Cosmos de Arag\'on (CEFCA). This pipeline corrects for bias, flat fielding, fringing, and illumination before calculating masks for cold/hot pixels, cosmic rays, and satellite traces. The different exposures are then calibrated astrometrically, PSF-homogenized, and coadded using the packages L.A.Cosmic \citep{vandokkum2001}, PSFEx \citep{bertin2011}, Scamp \citep{bertin2006}, SExtractor \citep{bertin1996}, and Swarp \citep{bertin2002}. We refer the reader to \cite{cenarro2018} for further details on the data reduction process.\\

We used spectroscopic redshifts to identify the cluster members in order to evaluate the total luminosity of the cluster (ICL plus the light enclosed by the galaxy members). We compiled a total of 1862 spectroscopic redshifts, gathered from the Sloan Digital Sky Survey Data Release 13 \citep[SDSS-IV DR13,][]{blanton2017,albareti2017}, the literature \citep{smith2009,mahajan2010,gavazzi2011} and NED\footnote{https://ned.ipac.caltech.edu/} (NASA/IPAC Extragalactic Database), having the precaution of rejecting those redshifts from NED which are derived from photometry or have poor quality. The SDSS spectroscopic catalog is 98\% complete up to a magnitude of $m_r\leq17.77$ (without considering the BOSS spectra), due to a limitation of the number of fibers in the spectrograph \citep{strauss2002,mahajan2010}. In some fields, objects with $m_r\leq17.77$ outnumber the amount of fibers available. As these regions are not revisited, the final completeness does not reach 100\%. At the redshift of the Coma cluster, 0.023, this magnitude depth is equivalent to an absolute magnitude of $M_r=-17.25$ mag. Using the luminosity function published by \cite{beijersbergen2002} and assuming that the photometric depth of our images is the same as that of the J-PLUS Early Data Release observations \citep{cenarro2018} ($m_r<=21.9$ mag), we calculate that we are underestimating the total luminosity of Coma by a $~9.3\%$ considering just spectroscopic information. Given that the photometry-based cluster membership algorithms are prone to contamination, especially at fainter magnitudes, we prefer to prioritize purity over completeness to obtain more 
accurate ICL fractions. We will analyze in Sect. \ref{sect_results} the impact of this underestimation on the final ICL fractions.\\

\section{Methodology} \label{sect_methodology}

The CHEFs Intracluster Light Estimator \citep[CICLE,][]{jimenez-teja2016,jimenez-teja2018} is an algorithm especially designed to estimate the ICL fraction accurately. It relies on the Chebyshev-Fourier functions \citep[CHEFs,][]{jimenez-teja2012} to model and remove out the luminous distribution of the galaxies. The detection and identification of the sources in the images is performed by Sextractor \citep{bertin1996}, while the modeling is made by the CHEFs. The CHEFs form mathematical bases, built in polar coordinates using Chebyshev polynomials and Fourier series, capable of fitting any smooth galaxy morphology with a few CHEF functions. CHEF bases are thus very compact and flexible, what makes them ideal to create robust models of the surface distribution of the galaxies and disentangle its light from the ICL. However, the brightest cluster galaxy (BCG) cannot be just modeled out with the CHEFs since its extended halo can be easily misidentified with ICL. We need a special analysis to identify the points where the transition from halo to ICL occurs. With this aim, we use a parameter called Minimum Principal Curvature \citep[MPC,][]{patrikalakis2002}, frequently used in differential geometry. The MPC is a characteristic of each point of a surface that highlights the change in the inclination of the surface at that point. CICLE's only required assumption is that the BCG extended halo and the ICL have different profiles (otherwise, they would be indistinguishable no matter what method was used), so that the limits of the BCG are defined by the curve of points with the highest MPC values.\\

Once we have the light from galaxies removed, an image containing just ICL and instrumental background is left. Here, we neglect cosmic optical background \citep[e.g.][]{zemcov2017} and other local atmospheric and solar system sources, which we pack under the designation of "instrumental". Although the ideal procedure to estimate this instrumental background would be to have a blank-sky observation off-target and measure the mean value from it, we use SExtractor with a special configuration instead, as follows. Our J-PLUS images are large enough to cover ICL-free regions in the clusters' outskirts, but we noticed that the instrumental background was not homogeneous across the entire field-of-view. For this reason, we preferred to use SExtractor to estimate these azimuthal fluctuations so they can be removed from the ICL estimations, thus avoiding the possible inaccuracies associated to incorrect extrapolations as much as possible. However, a special configuration is needed to guarantee that the background is not overestimated by misidentifying part of the ICL emission. We follow the prescription described in \cite{jimenez-teja2018}, where we select high values for the SExtractor background-related parameters, such as the  BACK\_SIZE or the  BACK\_FILTERSIZE. In short, SExtractor builds a continuous background map for the entire image interpolating a set of discrete background values estimated from the pixels not associated to any source. Each discrete background value corresponds to a certain region of the image, whose area is set by the BACK\_SIZE parameter. A typical value for BACK\_SIZE would be $\sim 16-32$, to include the small-scale fluctuations of the background. However, we set BACK\_SIZE = 512, a value unusually high to smear out the influence of high-valued pixels of the background and, therefore, reduce the influence of the ICL-related pixels. A second important SExtractor parameter that needs a special configuration is the BACK\_FILTERSIZE, which controls the size of the filter that is used to smooth the set of the discrete background values before interpolating them. Again, a narrow filter would keep small structures present in the background and would not smooth out the high ones, ICL-related values that could have passed the previous step. For this reason, we chose BACK\_FILTERSIZE = 5, with the clear strategy of avoiding any misidentification of ICL with background in spite of losing the fine details of the background. We refer the reader to \cite{jimenez-teja2018} for further details on the background estimation.\\

We also need an image with, exclusively, all the luminous components of the cluster to measure the total flux of the system. This image will thus be composed of ICL and the light from the cluster galaxy members. To identify these cluster members we follow a composite algorithm called the PEAK+GAP method \citep{owers2011}. The peak method performs, at first, a rough selection of possible candidates by examining the distribution in redshift of the galaxies. Once the peak in the redshift histogram related to the cluster is identified, we select as candidates all the galaxies lying in the region of the velocity distribution associated with this peak. As this subsample is likely contaminated by interlopers, a second and more sophisticated criterion is needed to identify them. The shifting gapper method \citep{fadda1996,girardi1996,boschin2006,owers2011} analyses the configuration of the candidates in the peculiar velocity-clustercentric distance space, split in radial bins. Those candidates with velocities similar to the mean velocity of the cluster and spatially closer to its center are chosen as cluster members. Galaxies with velocities too different from the mean of the bin are identified by an iterative algorithm that compares the velocities of the galaxies in pairs (previously sorted in ascending order), and are rejected as interlopers. \\

The CHEF models of the galaxies identified as members are reinserted into the ICL map to obtain the total luminosity map. The ICL fraction is then measured calculating the flux in sequentially larger areas, both in the ICL and the total luminosity maps. For the inner areas of the cluster, the natural contours of the ICL are used, while we fit ellipses for the outer regions. For consistency, the same contours are applied to both maps and the resulting radial flux profiles are compared to measure the ICL fraction. The radius up to which the ICL fraction can be measured is defined by the point where the ICL flux is minimum, right before starting to increase due to border effects of the image or other instrumental sources of error. The final error associated with the estimated ICL fraction comes from two sources: the photometric error and the error associated with the CICLE ability to disentangle the ICL from galaxy light. The first error is determined in a standard way, depending on the gain and rms of the observations as well as on the area and flux of the ICL and the BCGs \citep{bertin1996}. The second error is  estimated empirically, running CICLE on a set of ten mock images mimicking the geometry of the BCGs and the ICL of Coma, and the quality of the J-PLUS data. With this purpose, we generated a composite surface built with three exponential profiles with the same central positions, effective radii and surface brightness profiles  as the two dominant galaxies and the ICL of Coma. We then added noise with a dispersion such that the final image had the same signal-to-noise as the original J-PLUS observation. This last step is repeated ten times, to apply CICLE on these different, simulated images and take as final empirical error the average in quadrature of the ten individual errors. The final total error of our ICL measurements is the combination of the photometric and empirical errors, added in quadrature. \\

We refer the reader to \cite{jimenez-teja2012} for further information on the CHEFs, and to \cite{jimenez-teja2016} for a more detailed description of CICLE, including the background estimation, the cluster membership algorithm, and the estimation of the errors.\\ 

\section{CICLE results} \label{sect_results}

Before applying CICLE to our ten J-PLUS images of the Coma cluster, we built a mask to avoid the contamination from the multiple stars that appear in the field, especially the very bright, three stars located to the north and northwest of the BCGs, identified as HD 112887, HD 112886, and HD 112734. The CHEFs are bases of a mathematical space composed by squared-integrable, smooth functions. Therefore, they are unable to fit objects with discontinuities or sharp features such as stars, in particular those that are saturated. For this reason, the stars in the field are masked out using the image in the i band as reference. The SExtractor segmentation map corresponding to this image is used to visually identify the stars and mask out the pixels associated to them. We then enlarge these masked regions applying a uniform filter of 10  $\times$ 10 pixels, to guarantee that no light from these stars is misidentified as ICL. In special cases where SExtractor fails and deblends the star into several pieces, a circular mask is created manually. The final composite mask, joining the SExtractor-based and circular masks, is applied to all the filters so that we avoid any gradient bias in the ICL measurement and ensure that the possible differences in the ICL fraction are physical and not induced by different masking.\\

We processed the images from the ten filters described in Table \ref{tabla_observaciones_resultados} with CICLE. As a by-product of the analysis, we obtained an intensity-enhanced image of the ICL map in an intermediate step. After visual examination of these enhanced images (which is usually made as a CICLE internal sanity check), we noted that the observations in three of the filters had some artificial features. For the broad- and narrow-band filters {\it u} and {\it J0378} respectively, a ghost artifact surrounding the star right at the north of the BCG was clearly seen. In the case of the {\it z} band we noticed a strong fringing pattern. As these two features were very likely to pollute our ICL measurements, we decided to reject these three images and continue the analysis with the seven unaffected images left.\\

\begin{figure}
\centering
\includegraphics[width=\hsize]{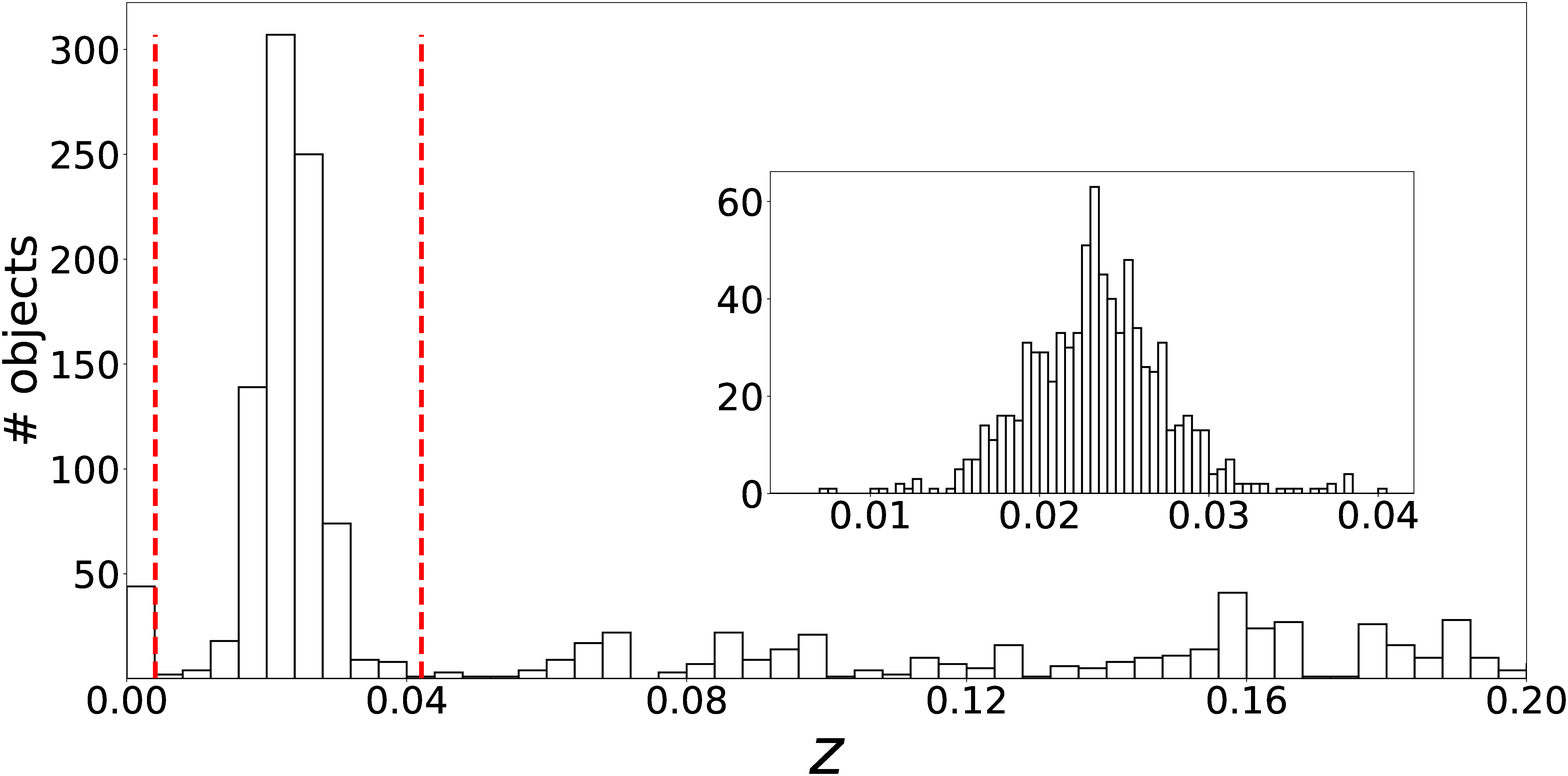}
\includegraphics[width=\hsize]{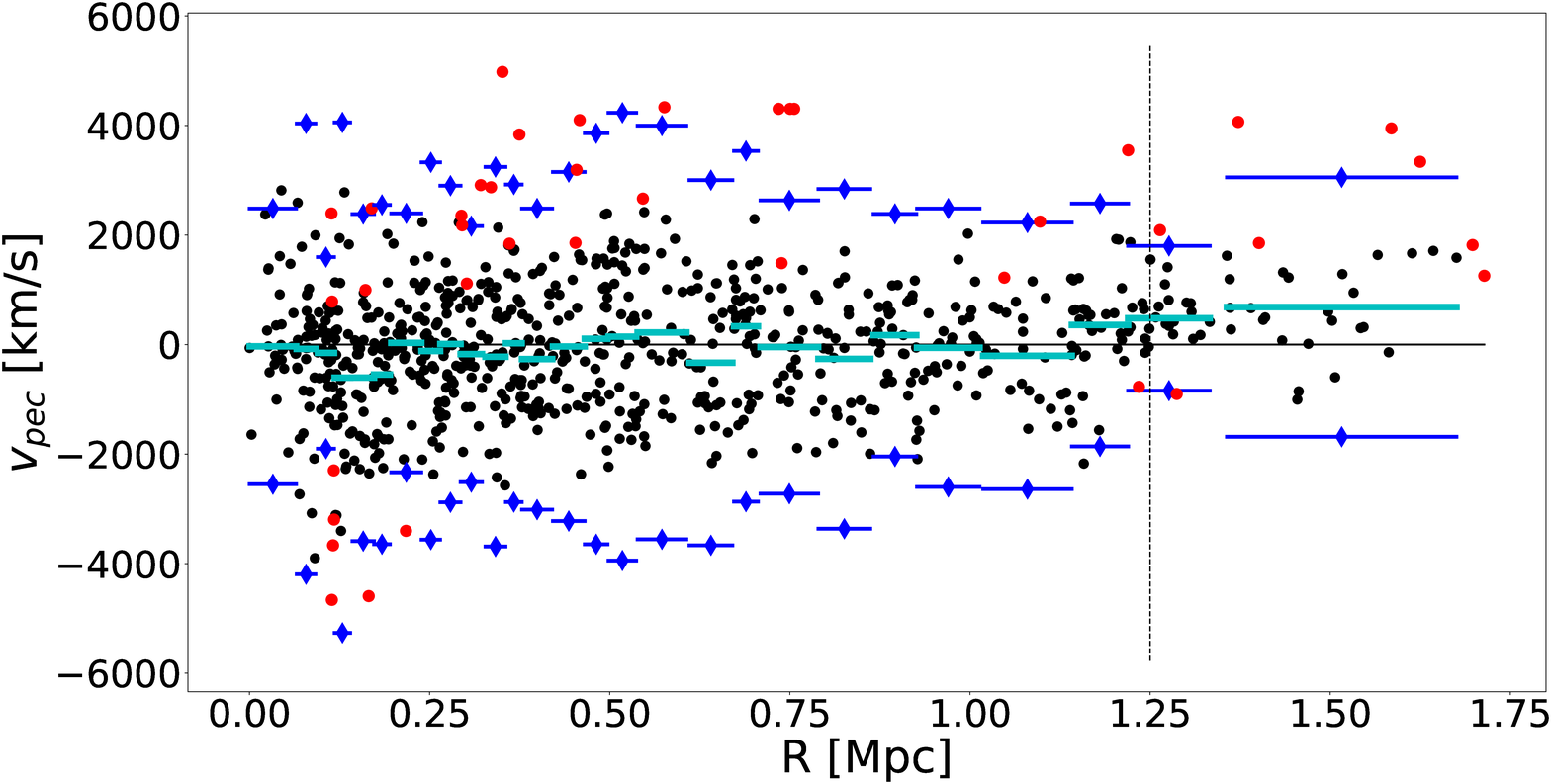}
\caption{Cluster membership for the Coma cluster with the PEAK+GAP algorithm. Top: redshift distribution of the objects in our spectroscopic catalog. Vertical red lines identify the redshift window with the possible member candidates, as obtained from the PEAK method. The inset displays a magnification of this window, with a smaller bin size for illustration purposes. Bottom: peculiar velocity-clustercentric distance plane from the GAP algorithm. Considered bona fide members are plotted with black points while rejected candidates are shown in red. Cyan horizontal lines with squared markers indicate the mean peculiar velocity on each bin, while blue lines with diamond markers correspond to the $3\sigma$ limit. Vertical black dotted line marks the radius up to which the gap algorithm is reliable, since the standard deviations systematically increase from this point. All the candidates outside this radius are discarded.}  \label{peak+gap}
\end{figure}

After applying the PEAK+GAP algorithm to our catalog of 1862 spectroscopic redshifts, we found that 782 were identified as cluster members (see Fig. \ref{peak+gap}). With their corresponding CHEF models and the ICL maps provided by CICLE, we estimated the ICL fractions listed in Table \ref{tabla_observaciones_resultados}. The projected area where the ICL fractions were measured is also indicated by the equivalent radius, assuming a circular shape.\\

Compared to the ICL fraction measured with CICLE for other clusters in previous works \citep{jimenez-teja2016, jimenez-teja2018}, we observe that the values estimated for Coma have larger errors. This is reasonable given that the previous analyses used imaging of exceptional quality and depth, as obtained by the CLASH \citep[Cluster Lensing and Supernovae Survey with Hubble,][]{postman} and FF programs \citep[Frontier Fields,][]{lotz2017} with the Hubble Space Telescope (HST). Moreover, the precision of CICLE depends on the geometry of the BCG+ICL system, yielding higher errors in the case where the two components have a similar shape (profile) or luminosity \citep{jimenez-teja2016}. In the case of the Coma cluster, we found that the two BCGs and the ICL had very similar fluxes and the signal-to-noise of the images was lower than that of the HST images, thus resulting on larger errors.
\\

As CICLE disentangles the ICL from the BCG by finding the points where the curvature changes most, the results are not biased by the effect of the PSF. As the PSF spreads the light of the sources, CICLE finds that the radii where the change of curvature occurs are larger in the case of PSF-convolved sources. In this way, CICLE always calculates total fluxes, so the estimated ICL fractions at different wavelengths can be directly compared, without the need of applying any correction. In order to confirm this statement, we calculated the PSFs of the seven J-PLUS images analyzed and created mock images with the same observational characteristics as the original BCG+ICL surface (surface brightness, effective radius, signal-to-noise, etc). The test consisted in applying CICLE to the simulated images before and after convolving with each corresponding PSF. Then, the flux of the ICL estimated by CICLE was compared to the real flux of the simulated profiles. Results showed that CICLE systematically found the transition from the BCG to the ICL dominated region at larger radii in the case of the PSF-convolved images. In relation with the unconvolved images, the error in the estimation of the ICL flux in the PSF-convolved images was larger for the {\it J0410, J0515,} and {\it g} filters, smaller for the {\it J0430}, and nearly unchanged for the {\it J0395, r,} and {\it i} bands. Thus, no correlation is observed between the estimated ICL fractions with wavelength and no artificial bias should be introduced when the ICL fractions of different filters are compared.\\

\section{Discussion} \label{sect_discussion}

We will analyze the implications of the results by: a) comparing the ICL fractions obtained for Coma with those of the clusters studied in the previous works with HST data, and b) studying the morphology of the luminous distribution of the ICL, as in \cite{adami2005}.

\subsection{Analysis of the ICL fraction: the blue excess}

In \cite{jimenez-teja2018} we analyzed a sample of eleven CLASH and FF galaxy clusters in the redshift range $0.18<z<0.54$ with their dynamical stages clearly stated by different indicators. Measuring their ICL fractions in three different HST optical filters (F435W, F606W, F814W) we found that the contribution of the ICL to the total luminosity of the cluster was similar and nearly wavelength independent for the subsample of relaxed clusters (with ICL fractions varying in the interval $\sim 2\%-11\%$), which would be expected if the stellar populations in the ICL and the cluster galaxies were similar and might just be equally evolving passively. However, for the subsample of merging clusters we observed higher ICL fraction  values in general in the three filters ($\sim 7-23\%$), which was significantly accentuated in the case of the F606W band. Filter F606W was identified as the one mainly comprising the emission peaks of F- and A-type stars at this redshift range. Given that \cite{morishita2017} identified a non-negligible amount of A- or earlier-type stars contributing significantly to the ICL mass budget in the FF clusters (between 5\% and 10\%), we suggested that the ICL fraction excess in the F606W band could be originated by these bluer stars that might have been extracted from the outskirts of the infalling galaxies and groups to the ICL during the merging event. This excess ICL fraction observed in a particular wavelength band in merging clusters opened a "clean" window to determine the dynamical state of galaxy clusters based solely on the spectrum of the ICL fraction. If our suggested hypothesis for this effect is correct, we should expect an ICL excess at about 3800\AA--4800\AA~ for a cluster dynamically active at redshift near zero, i.e., reacting to a recent merging, and this is what our observations suggest for the Coma cluster as described below.\\

In Coma we find relatively high ICL fractions (between $\sim 7\%$ and $\sim 21\%$, depending on the filter used), which are consistent with previous works \citep{melnick1977,thuan1977,mattila1977,bernstein1995,adami2005}. These might be indicative of a non-relaxed state of the cluster according to the values found for higher-redshift systems in \cite{jimenez-teja2018}. In fact, compared to other nearby galaxy clusters such as Fornax \citep{iodice2017} or Virgo \citep{mihos2017}, Coma shows more discrepancies with the former. Fornax is believed to be a dynamically evolved, non-merging cluster, which is translated into an ICL color distribution representing well that of its member galaxies and also a low ICL fraction (less than ~5\% in the g-band).  Its ICL is believed to be formed from stars and globular clusters stripped out from the outskirsts of galaxies, as we suggest for the case of Coma. Virgo shows many indications of being in an unrelaxed stage, such as the substructure observed in both its spatial and kinematic distributions, or the variety of morphological types displayed by its galaxy members. As Coma, Virgo presents some ICL tidal features that are bluer than the inner regions of its massive ellipticals, although this effect is more pronounced in the case of Coma. Although measured with a different technique, it has an estimated ICL fraction of $\sim 7\%-15\%$ in the V-band, comparable to the range of $\sim 7\% - 21\%$ spanned by Coma (albeit somewhat smaller) and consistent with the sample of merging clusters analyzed in \cite{jimenez-teja2018}.\\

\begin{figure}
\centering
\includegraphics[width=\hsize]{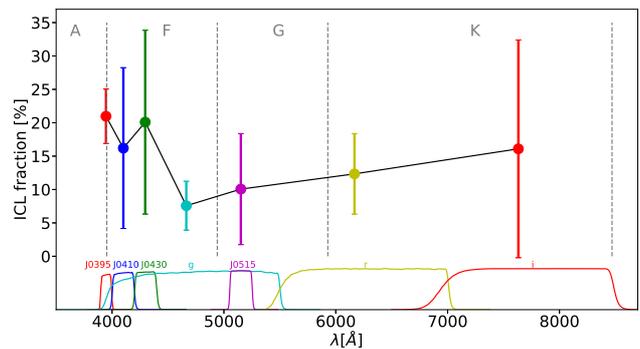}
\caption{ICL fractions for Coma as a function of the wavelength. Transmission curves of the seven J-PLUS filters analyzed are shown, with the same color code as the ICL fractions. Vertical gray lines delimit the intervals were the emission peaks of each stellar spectral type are contained, as indicated at the top of each region.}  \label{ICLfraction-lambda}
\end{figure}

However, as previously described, the most characteristic feature to confirm its dynamical stage is the possible excess in the filter (or filters) to which the flux of the F- and A-type stars contributes most. At the redshift of the Coma cluster, $z\sim0.023$, if the system were in an active phase, due, e.g., to a recent merger, we would expect an ICL fraction peak at about 4400\AA~$\pm\sim$ 500\AA. In Fig. \ref{ICLfraction-lambda} we present the ICL fraction obtained for each filter. Transmission curves of the filters are also plotted at the bottom to show how the narrow and broad filters overlap, so their associated ICL fractions are not completely independent. As in \cite{jimenez-teja2018}, we use the emission peaks of the different stellar types at the redshift of the cluster to split the covered wavelength range into four regions that can be compared with the transmission curves. We again identify a peak in the ICL fraction measured for the filters that mainly comprise the emission peaks of F- and A-type stars at the redshift of the Coma cluster, namely the bluer narrow filters {\it J0395, J0410,} and {\it J0430}. Although the {\it g} band also encompasses the emission peaks of the F-type stellar population, its contribution seems to be smoothed out by the width of the filter. Moreover, it should be noted that the filter partial overlap makes the ICL fraction excess in the blue filters measured here a conservative lower limit estimate. From Fig \ref{ICLfraction-lambda} it is clear that the {\it g} filter overlap with the bluer narrow bands should reduce the gradient intensity. \\

A simple test to estimate the significance for a change in the ICL fractions of the bluer, narrow bands {\it J0395, J0410,} and {\it J0430} compared to those of the redder filters {\it g, J0515, r,} and {\it i} is comparing their error-weighted averages. Doing so we get that gradient observed in the ICL fraction is significant at $2.4\sigma$. A more sophisticated statistical analysis is testing the null hypothesis that the ICL fraction dependence on the wavelength is flat. We this aim, we randomly generate different ICL fraction values within the error bars for each filter, and apply a Kolmogorov-Smirnov test to each realization of the ICL fractions. After $10^6$ realizations we obtain that the null hypothesis is rejected (and thus there exists a wavelength dependence/gradient in the ICL fraction) in 99.99\% of the cases with a significance of $2.6\sigma$. This gradient is consistent with what we found for our higher redshift subsample of merging clusters in \cite{jimenez-teja2018}. \\

Analyses of the population of intracluster globular cluster (IGC) in the Coma cluster core \citep{peng2011,cho2016}, revealed that the IGC color distribution is bimodal, with blue, metal-poor IGCs outnumbering red, metal-rich IGCs. In spite of observing a bluer ICL, they find that the fraction of red IGCs is non-negligible, representing 20\% of the IGCs in the cluster core. Our study of the ICL fraction with narrow filters in the Coma cluster indeed confirms that the ICL is bluer than the galaxy members of the cluster, as in previous works \citep{mattila1977,thuan1977}, while at the same time we see a significant amount of red ICL stars.\\

This is consistent with a scenario where part of the ICL is composed of bluer, younger/lower-metallicity stars stripped from the outskirts of infalling galaxies or dwarfs disrupted during past/ongoing merging events \citep{peng2011,jimenez-teja2018}. Indeed, in their study of 142 early-type galaxy members of Coma (mainly drawn from the core), \cite{denbrok2011} found that almost all of them had negative radial color gradients, indicating that stars in the outskirts of the galaxies are bluer than those closer to the center. This is also in agreement with recent integral field unit results of nearby galaxies with CALIFA \citep{gonzalez-delgado2015}. Although \cite{denbrok2011} did not observe any dependency of the steepness of the color gradient with the radial distance to the center of the cluster (which they noted that could be a selection effect), a comparison with galaxies in a lower-density environment as Fornax revealed that the latter had much stronger metallicity gradients. Moreover, the size of the outer disks of lenticular and spiral galaxies in the center of the Coma cluster are a $\sim 30-82\%$ smaller than their counterparts in lower dense environments (e.g. \cite{weinzirl2014} and references therein). These facts would back up not only the hypothesis that these galaxies could have already lost their outer stellar layers (as already pointed out by \cite{thuan1977}), causing a bluer ICL and flattening the color gradient of Coma galaxies, but also the idea that the merging process in the core of Coma started more than a crossing time ago, and is in a very advanced (however, ongoing) stage. Assuming an average age for an A/F-type star of $\sim 2.3\times 10^9$ years using a Salpeter IMF from $1.1 - 2.5 M_{\odot}$, and also assuming that they were extracted halfway through their lifetimes, the presence of this ICL fraction excess would indicate that the merger happened probably less than $\sim$ 1.2$\times 10^9$ years ago.\\

\begin{figure*}
\centering
\centerline{\includegraphics[width=9cm]{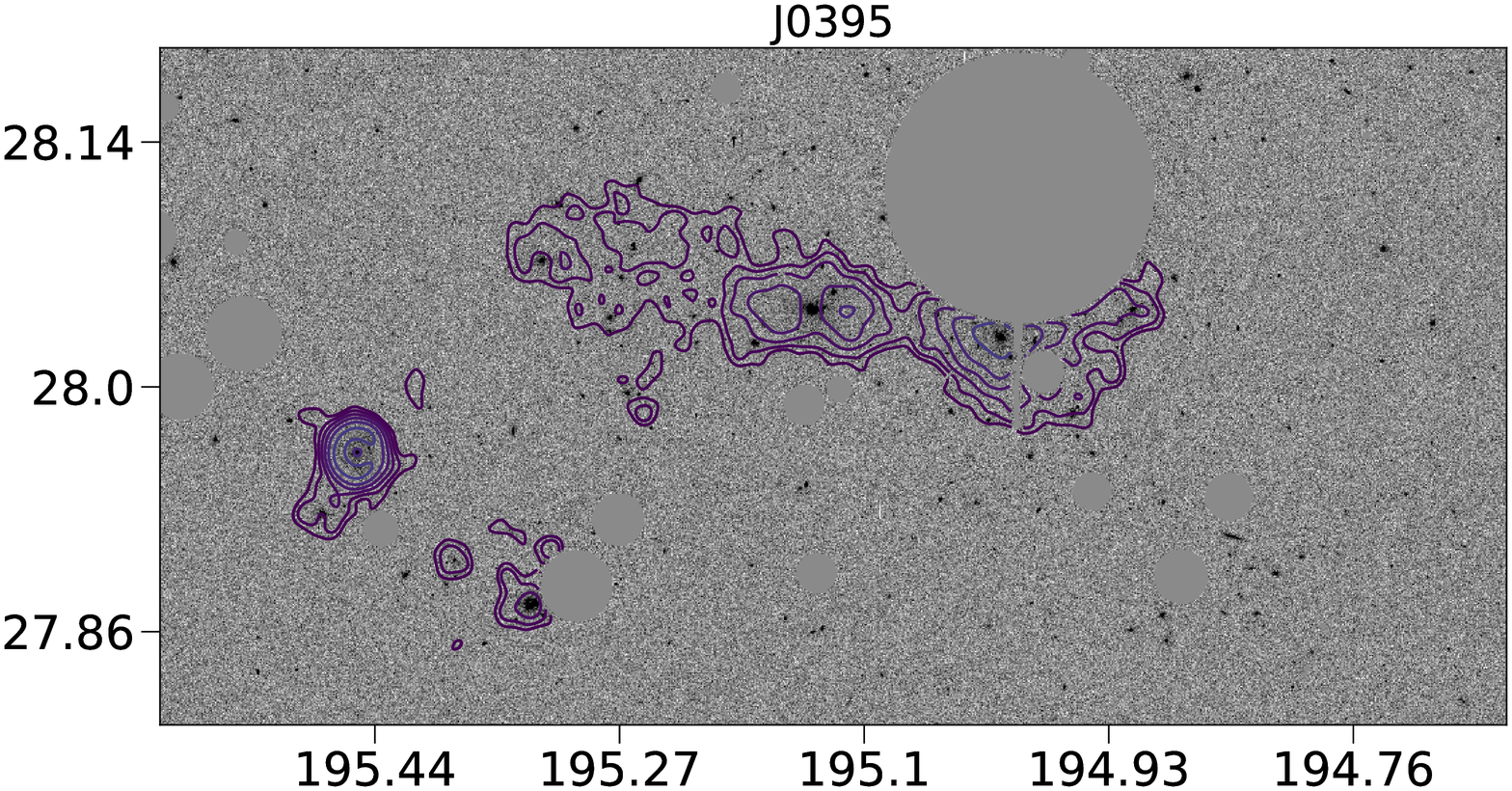}\includegraphics[width=9cm]{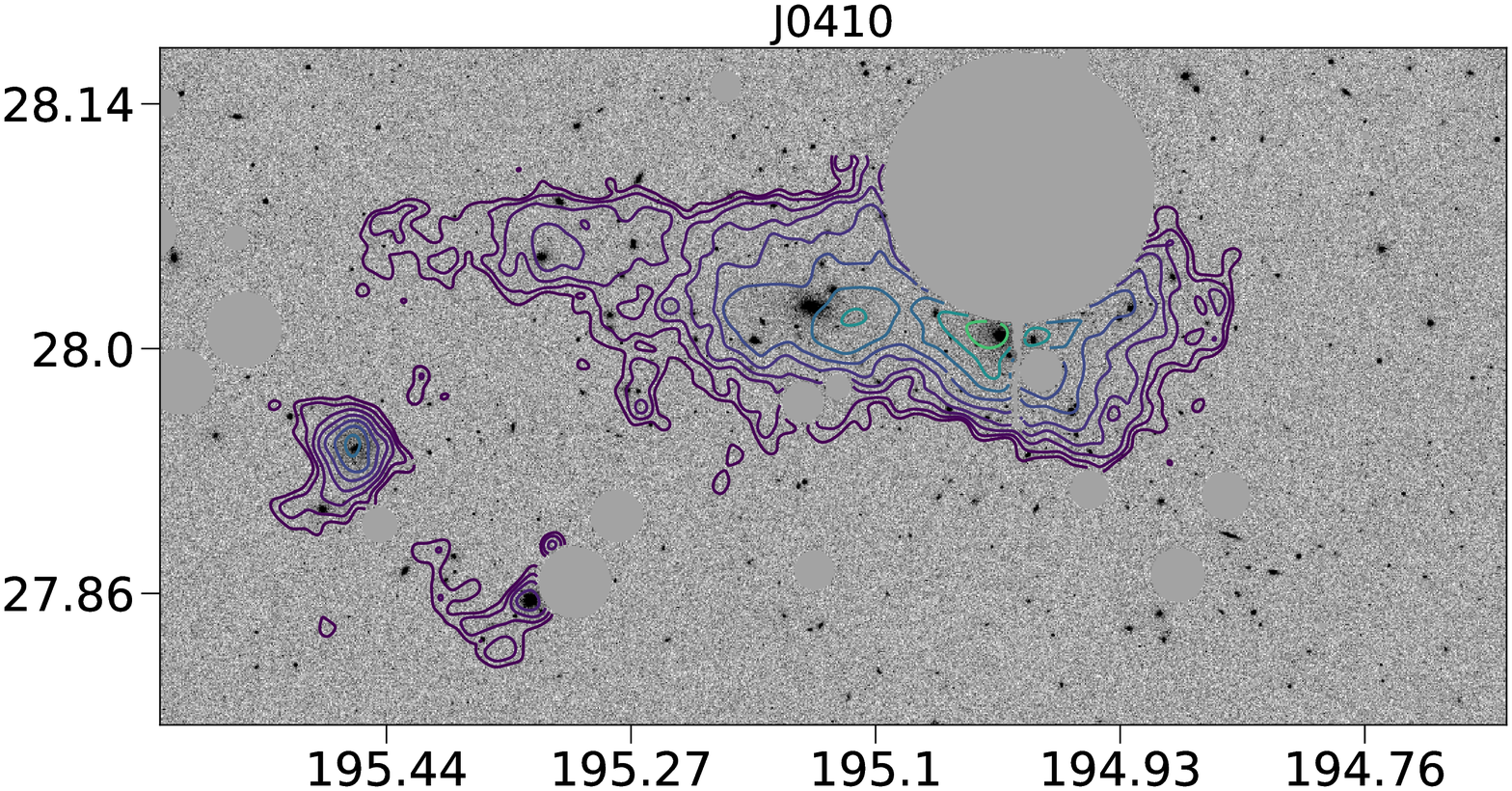}}
\centerline{\includegraphics[width=9cm]{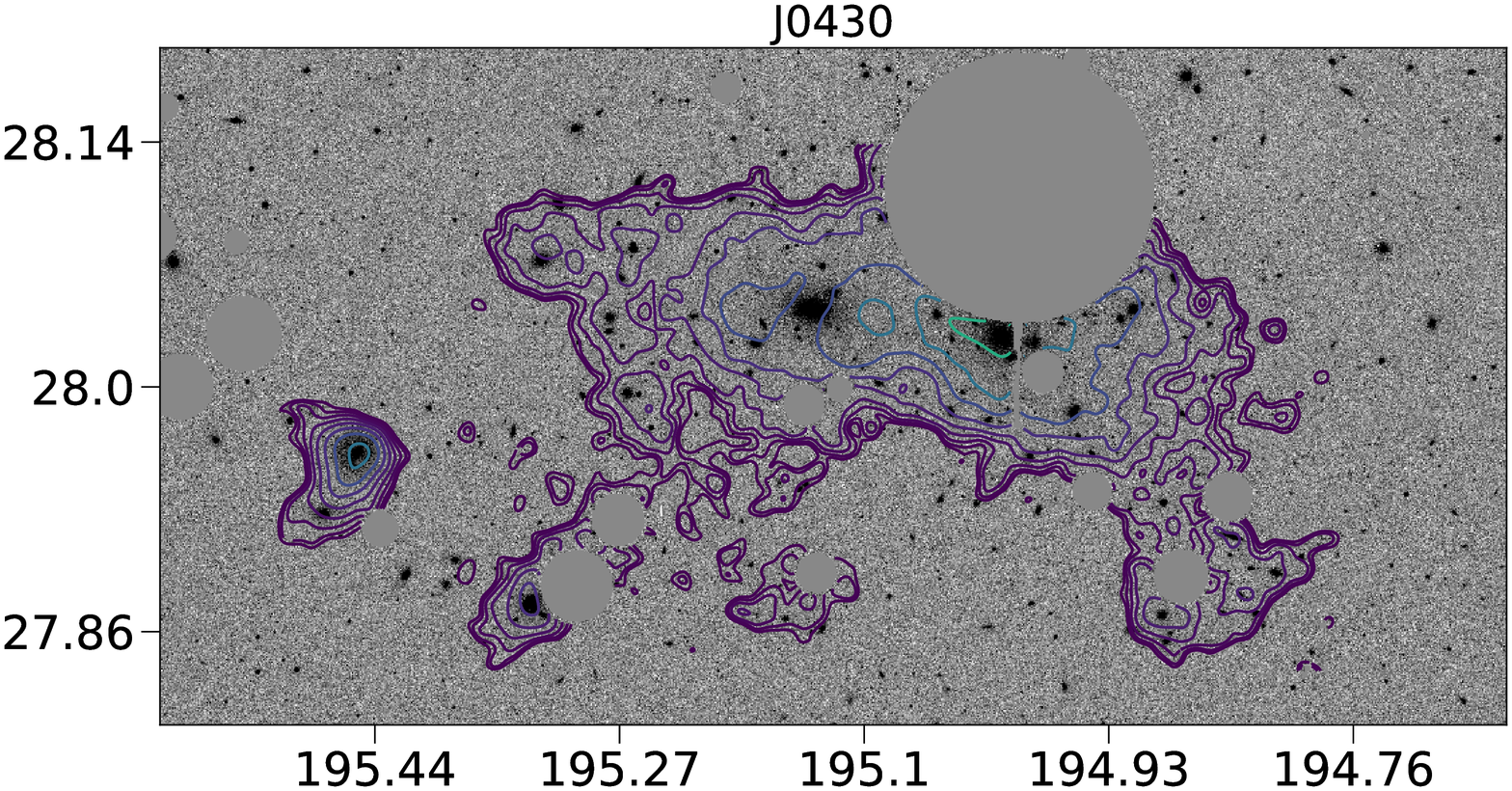}\includegraphics[width=9cm]{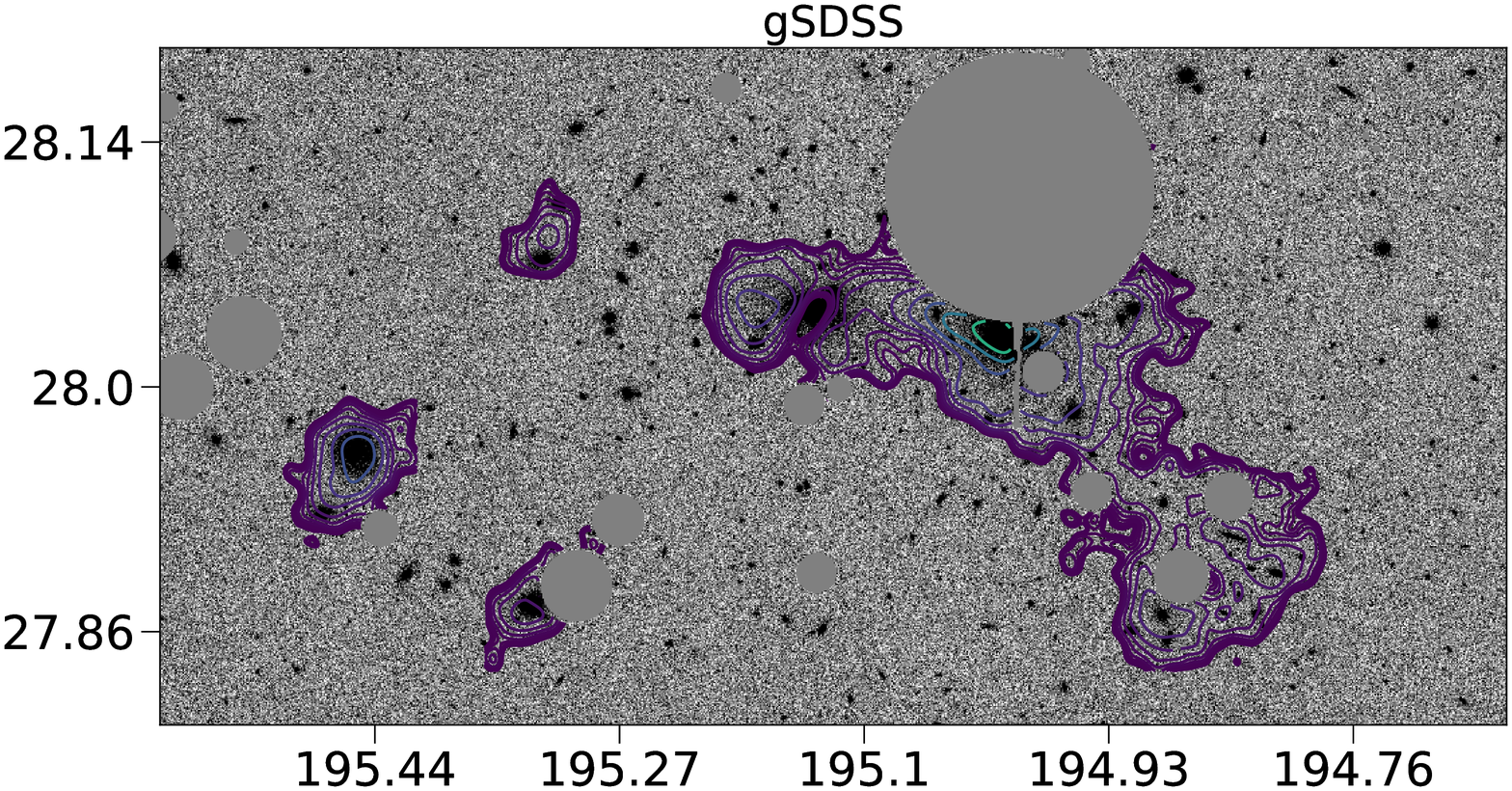}}
\centerline{\includegraphics[width=9cm]{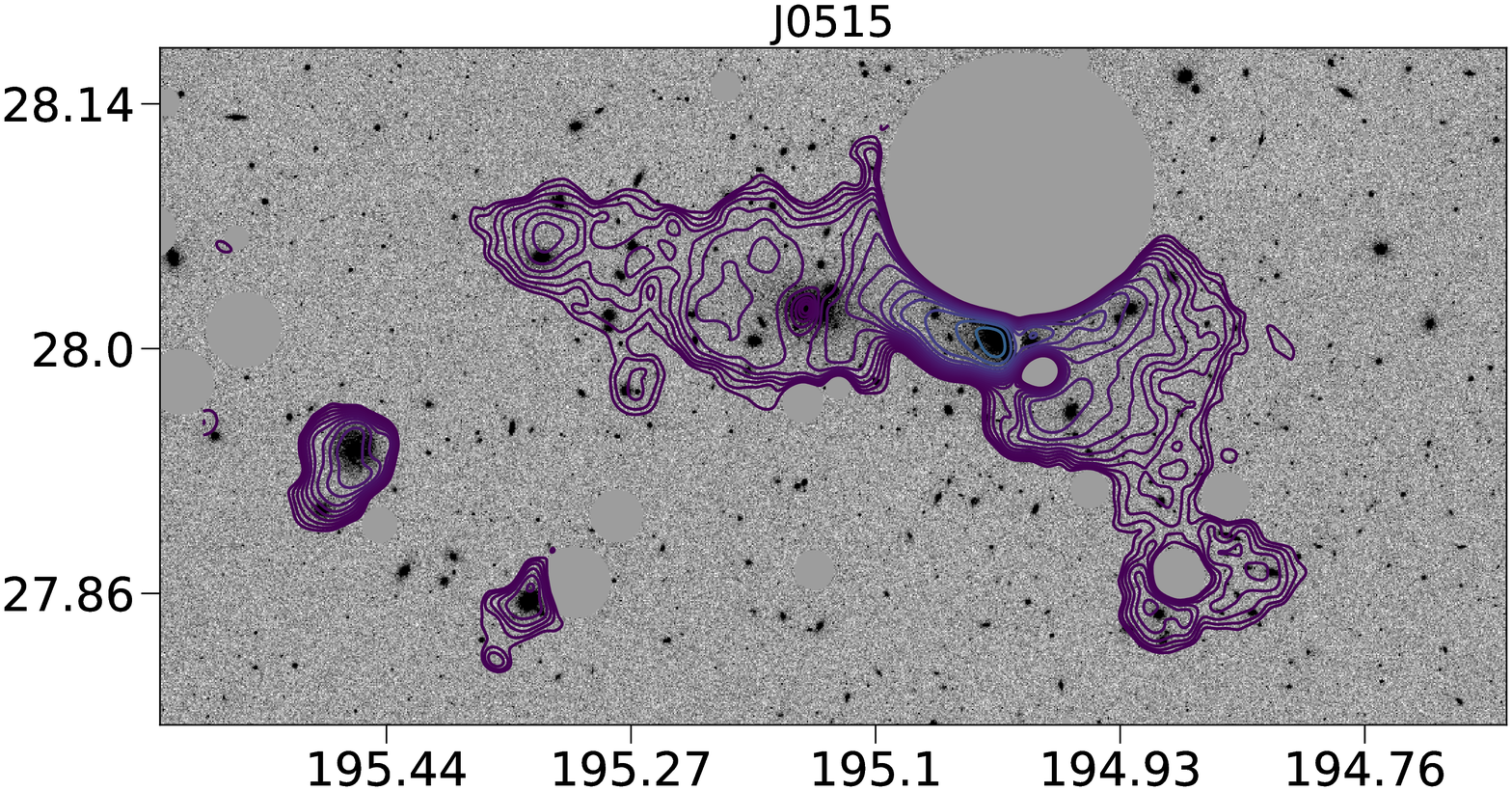}\includegraphics[width=9cm]{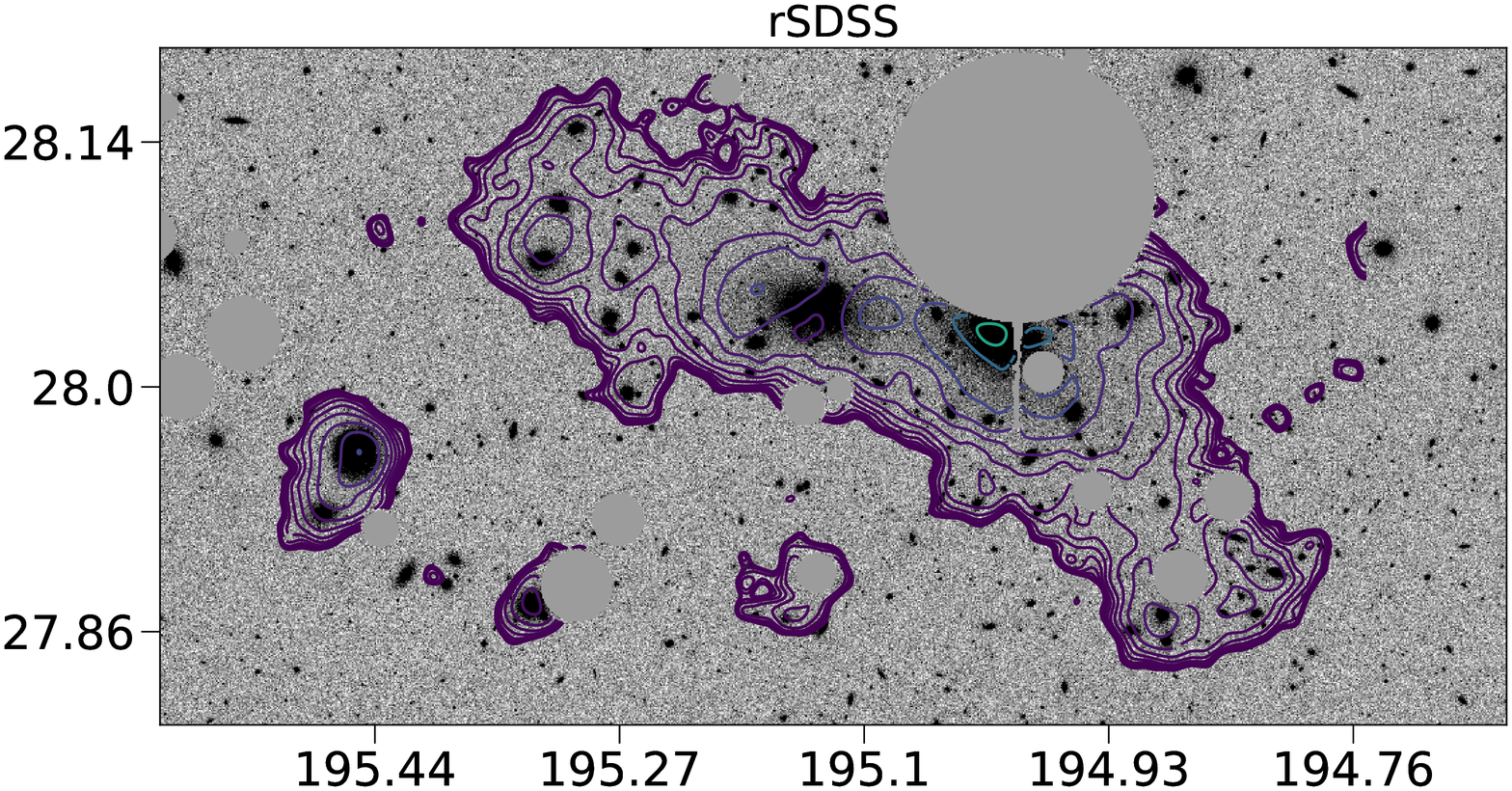}}
\centerline{\includegraphics[width=9cm]{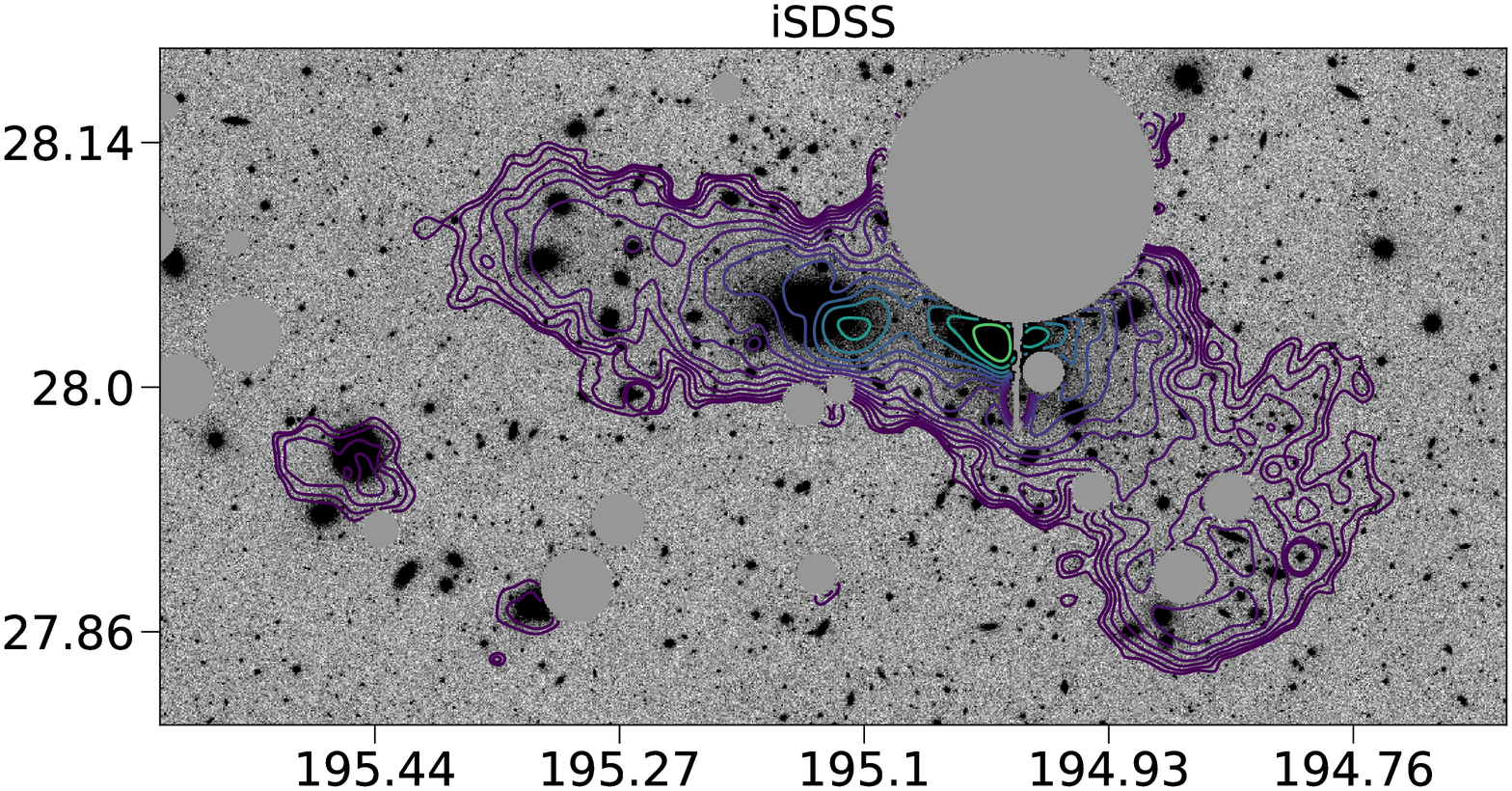}\includegraphics[width=9cm]{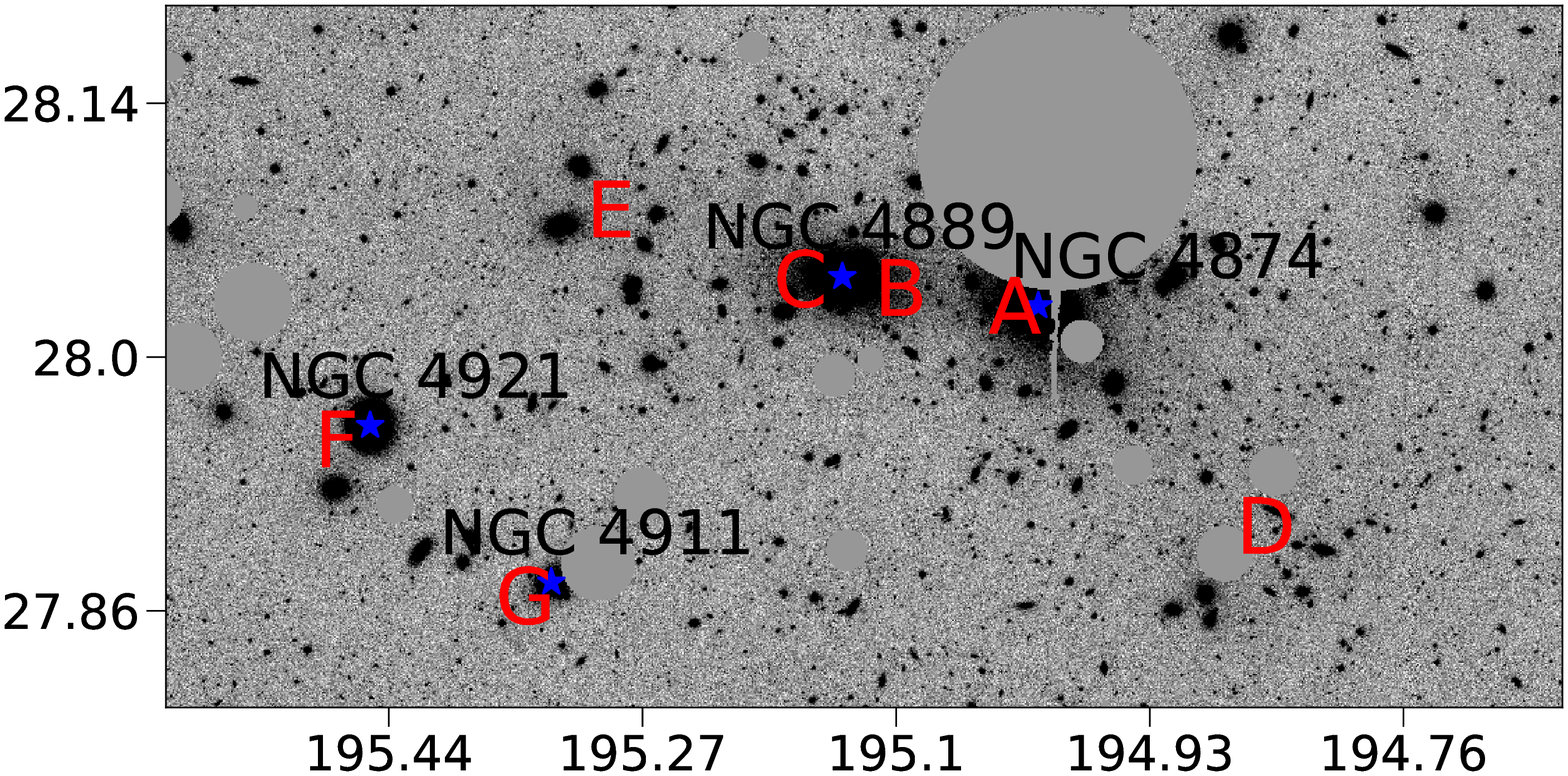}}
\caption{Contour maps of the ICL in the central region of the Coma clusters superimposed on the original data, and guiding image to identify the ICL substructure (bottom right). Original images have the stars and spikes masked out. We have smoothed the surface distribution of the ICL with a Gaussian of 17.6 arcsec typical deviation to draw the contours. Masked out pixels were not considered neither for the estimation of the ICL fraction nor the calculation of contours.  Guiding image at the bottom right shows the original image in the {\it i} band with the main galaxies (blue stars) and  ICL clumps (red letters) labeled. North is up and East to the left.}     \label{ICLcontours}
\end{figure*}

\subsection{Morphology of the ICL}

In Fig. \ref{ICLcontours} we show spatial distribution of the ICL in the core of the Coma clusters for the seven J-PLUS filters analyzed. We observe that the diffuse light is mainly spread over a common envelope containing the two giant galaxies NGC 4874 (right below the masked, very bright star with spikes) and NGC 4889 (east of NGC 4874). The ICL distribution is clearly elongated toward east, as reported by many authors in the literature \citep{kormendy1974,mattila1977,thuan1977,adami2005}. This lack of symmetry confirms again that Coma is far from being relaxed. For wavelengths redder than the {\it J0430} band, we see a clear arc-shaped extension of diffuse light spreading to the southwest from NGC4874 \citep[as in][]{welch1971,kormendy1974,mattila1977,melnick1977}, which is associated with a change in the ICL position angle, from $\sim 80$ degrees in the narrow blue filters to $\sim 55-65$ degrees in redder bands. This implies that the red stars in the ICL follow closely the overall distribution of the galaxy members, since the cluster major axis is estimated to be at a position angle of 64-65 degrees \citep{rood1972,thuan1977}. \\

Previous works reported that the ICL southeast edge is very sharp (specially south of NGC 4889) compared to the northwest bridge, whose intensity drops smoothly \citep{kormendy1974,melnick1977}. Apart from its morphology, the dynamical activity in Coma is reflected in the paucity of smoothness and homogeneity of the ICL, where we can observe substructures. From the bottom right panel in Fig. \ref{ICLcontours} we can conclude that the diffuse light is mainly concentrated around NGC 4874 (clump A) at every wavelength (consistently with the distribution found by \cite{adami2005}), although we cannot be sure that the ICL emission peak and the centroid of the giant elliptical are completely aligned due to the proximity of the bright star that might be altering the isocontours. We do see the absence of an emission peak centered on NGC 4889. Instead, the diffuse light is preferentially concentrated on two clumps, east and west of this giant galaxy, evident in all the filters although most intensely in the bluer narrow bands (marked in Fig. \ref{ICLcontours} as B and C). Although a bit displaced with respect to our measurements, the existence of an ICL clump in between the two D galaxies (B) was already noted by \cite{adami2005} (as well as several other substructures north to NGC 4874 that we are unable to confirm because they lay over our masked area). Given the expected strong tidal forces occurring between the two giant galaxies, it is not surprising that this region (B) is observed to be ICL rich. Stripping of the stars in the outskirts of the massive galaxies and the total disruption of dwarfs would be more efficient mechanisms in this region.  \\

To our knowledge, the clump C, east of NGC 4889, was never reported. Comparing its location with the likely orbits of the two dominant galaxies during the merger (Fig. 2 of \cite{gerhard2007}), we suggest that it might have been formed from the surroundings of NGC 4874 and its associated group during its travel to the bottom of the potential well. A dynamical study of the  velocity distribution of planetary nebulae in this region could check the association with NGC4874, testing this scenario.\\

Apart from these three main clumps of ICL closer to the center of Coma, we observed at least three other different structures in Fig. \ref{ICLcontours}: the reddish arc-shaped ridge spreading toward southwest from NGC 4874 (D); the more subtle, elongated extension of light east of NGC 4889 (E); and the two very blue clumps associated with the spiral galaxies NGC 4921 and NGC 4911, southeast from NGC 4889 (F and G respectively).\\

Both D and E substructures have been previously observed by other authors \citep{welch1971,kormendy1974,melnick1977, thuan1977}. They are very irregular according to their isocontours: multiple small peaks of ICL are distributed along these regions, apparently without a dominant clump. The slope of the diffuse light distribution is small in both areas, compared to the distribution in the A, B, and C clumps of the cluster core. A significant difference is that the SW extension D (ridge) does not appear in the bluest filters while the E substructure is visible in all bands. The lack of bluer ICL in D may suggest that this is an earlier region affected by the merger, where most of the the stripped A- and F-type stars are already dead. The morphological density contour maps calculated by \cite{adami2008} reveal density clumps in the E region for all morphological types, although strongly dominated by early type galaxies. With a bit less intensity, the D substructure is also primarily dominated by early type galaxies, although it shows a non-negligible contribution from irregulars too. However, these density maps show a clear deficit of spirals and starbursts in this region, which would explain its lack of luminosity in the bluest filters. \\

Early works on the ICL of Coma had already reported the presence of the clumps F and G \citep[e.g. ][]{melnick1977}. Neither of them seem to be connected to the main envelope of ICL surrounding the giant ellipticals of the center, at least down to our detection limit. Just in the {\it J0430} band, clump G appears weakly connected to the main ICL structure, but with low significance. Both of them are very blue whereas they appear much weaker in the red bands, which is not surprising given that these regions show a higher density of irregular and starbust galaxies than the surroundings \citep{adami2008}. In particular, the F clump is as blue as the two main concentrations of ICL in the core (A and B). According to \cite{adami2005b}, the two spirals NGC 4921 and NGC 4911 belong to two dynamically independent groups that could be falling into the cluster following tangential orbits. This is in agreement with the fact that clumps F and G show no significant indication of being linked at any wavelength, according to our ICL contour maps.\\

\section{Conclusions} \label{sect_conclusions}

In \cite{jimenez-teja2018} we showed that the ICL fraction distribution across different wavelengths could be used as an estimator of the dynamical state of massive systems using HST data of a sample of galaxy clusters distributed in the redshift range $0.18<z<0.54$. Our new analysis of the ICL fraction of Coma not only confirms that it is a dynamically active system, but also that the ICL fraction can be an indicator of the dynamic stage for nearby clusters, given an appropriate selection of the filters. The J-PLUS filter system has been proved to be an appropriate set to study the ICL fraction distribution across wavelengths in nearby clusters, given its combination of narrow- and broad-band filters. In fact, we must remark that using the traditional SDSS filter system solely we might not significantly detect the ICL fraction excess, characteristic of dynamical activity.\\

With Coma as a case study, we have also tested the capabilities and the quality of the J-PLUS data to analyze the diffuse light in galaxy clusters in terms of depth and precision. Despite the larger errors due to the special geometry of the Coma BCG+ICL system and the lower signal-to-noise of the J-PLUS images (compared to that of the HST observations), we proved that we can obtain reliable measurements of the ICL fraction with CICLE applied on J-PLUS data. Estimated ICL fractions range from $\sim 7\%$ to $\sim 21\%$, consistent with previous values reported in the literature for similar wavelengths (usually green and red filters). We also extended the traditional study of the diffuse light in Coma to bluer bands, revealing the characteristic excess in the ICL fraction measured in the filters that comprises the emission peaks of F- and A-type stars, typical of merging clusters \citep{jimenez-teja2018}.\\

We also took the advantage of the multicolor J-PLUS system to analyze the varying morphology of the ICL at different wavelengths (see Fig. \ref{ICLcontours}). The complex, elongated, asymmetric and clumpy structure of the ICL confirms that Coma is far from a virialized state. CICLE results confirmed several features previously reported in the literature, like the main concentration around the NGC 4874 (A), the clump in between the two BCGs (B), the elongated extension toward east (E), the reddish arc-shaped feature southwest from NGC 4874 (D), and the blue concentrations associated with the spirals NGC 4921 and NGC 4911 (F and G, respectively) \citep[e.g.][]{melnick1977,thuan1977,adami2005}. We also found a new feature: a multicolored peak east of NGC 4889 (C), visible through all our wavelength range. The distribution and colors of all these features represent independent evidence that confirms the dynamical scenario proposed by several authors in previous works for the Coma cluster \citep[e.g.][]{colless1996,adami2005b,gerhard2007}.\\

Although the obtained results are consistent with those found in the literature, in the future we aim to complement our study with new J-PLUS observations. Higher signal-to-noise images and the study of the filters that were not included in the present study, either because they were not observed or the images presented some kind of artifact, will be essential to reduce the errors of the estimated ICL fractions and increase the significance of the ICL fraction excess. A study of the stellar populations in different regions of the ICL of Coma will be described in a future paper, using the J-PLUS photometry as a low-resolution spectra. \\

\begin{acknowledgements}

We gratefully acknowledge the computational support of Dr. Fernando Roig. Y. J-T. also acknowledges financial support from the Funda\c{c}\~ao Carlos Chagas Filho de Amparo \`a Pesquisa do Estado do Rio de Janeiro - FAPERJ (fellowship Nota 10, PDR-10) through grant E-26/202.835/2016. R.A.D. acknowledges support from the Conselho Nacional de Desenvolvimento Cient\'ifico e Tecnol\'ogico - CNPq through BP grant  312307/2015-2,  and the Financiadora de Estudos e Projetos - FINEP grants  REF. 1217/13 - 01.13.0279.00 and REF 0859/10 - 01.10.0663.00 for hardware support for the J-PLUS project through the National Observatory of Brazil and CBPF. Both Y. J-T. and R.A.D. also acknowledge support from the Spanish National Research Council - CSIC (I-COOP+ 2016 program) through grant COOPB20263, and the Spanish Ministry of Economy, Industry, and Competitiveness - MINECO through grants AYA2013-48623-C2-1-P and AYA2016-81065-C2-1-P. R.L.O. was partially supported by the Brazilian agency CNPq (PDE 200289/2017-9, Universal Grants 459553/2014-3, PQ 302037/2015-2). J. A. H.J. thanks to Brazilian  institution CNPq for financial support through  postdoctoral fellowship (project 150237/2017-0). Funding for the J-PLUS Project has been provided by the Governments of Spain and Aragon through the Fondo de Inversiones de Teruel, the Spanish Ministry of Economy and Competitiveness (MINECO; under grants AYA2015-66211-C2-1-P, AYA2015-66211-C2-2, AYA2012-30789, and ICTS-2009-14), and European FEDER funding (FCDD10-4E-867, FCDD13-4E-2685). We thank the OAJ Data Processing and Archiving Unit (UPAD) for reducing and calibrating the OAJ data used in this work. This research has made use of the VizieR catalogue access tool, CDS, Strasbourg, France. Funding for the Sloan Digital Sky Survey IV has been provided by the Alfred P. Sloan Foundation, the U.S. Department of Energy Office of Science, and the Participating Institutions. SDSS-IV acknowledges support and resources from the Center for High-Performance Computing at the University of Utah. The SDSS web site is www.sdss.org. SDSS-IV is managed by the Astrophysical Research Consortium for the Participating Institutions of the SDSS Collaboration including the Brazilian Participation Group, the Carnegie Institution for Science, Carnegie Mellon University, the Chilean Participation Group, the French Participation Group, Harvard-Smithsonian Center for Astrophysics, Instituto de Astrof\'isica de Canarias, The Johns Hopkins University, Kavli Institute for the Physics and Mathematics of the Universe (IPMU) / University of Tokyo, Lawrence Berkeley National Laboratory, Leibniz Institut f\"ur Astrophysik Potsdam (AIP),  Max-Planck-Institut f\"ur Astronomie (MPIA Heidelberg), Max-Planck-Institut f\"ur Astrophysik (MPA Garching), Max-Planck-Institut f\"ur Extraterrestrische Physik (MPE), National Astronomical Observatories of China,New Mexico State University, New York University, University of Notre Dame,Observat\'ario Nacional / MCTI, The Ohio StateUniversity, Pennsylvania State University, ShanghaiAstronomical Observatory, United Kingdom Participation Group,Universidad Nacional Aut\'onoma de M\'exico,University of Arizona, University of Colorado Boulder, University of Oxford, University of Portsmouth, University of Utah, University of Virginia,University of Washington, University of Wisconsin, Vanderbilt University, and Yale University.
      
\end{acknowledgements}

%
\end{document}